# SmB$_6$ Photoemission: Past and Present

Jonathan D. DENLINGER*, James W. ALLEN[1], Jeong-Soo KANG[2], Kai SUN[1], Byung-Il MIN[3], Dae-Jeong KIM[4] and Zachary FISK[4]

*Advanced Light Source, Lawrence Berkeley National Laboratory, Berkeley, CA 94720, USA*
[1]*Dept. of Physics, Randall Laboratory, University of Michigan, Ann Arbor, MI 48109, USA*
[2]*Dept. of Physics, The Catholic University of Korea, Bucheon 420-743, Korea*
[3]*Dept. of Physics, POSTECH, Pohang 790-784, Korea*
[4]*Dept. of Physics and Astronomy, University of California at Irvine, Irvine, CA 92697, USA*

E-mail: JDDenlinger@lbl.gov

Recent renewed interest in the mixed valent insulator SmB$_6$ comes from topological theory predictions and surface transport measurements of possible in-gap surface states whose existence is most directly probed by angle-resolved photoemission spectroscopy (ARPES). Early photoemission leading up to a recent flurry of ARPES studies of in-gap states is reviewed. Conflicting interpretations about the nature of the Sm 4*f*-5*d* hybridization gap and observed X-point bands between the *f*-states and the Fermi level are critically assessed using the important tools of photon polarization and spatial dependence which also provide additional insight into the origin of the more ambiguous Γ-point in-gap states.

## 1. Introduction

SmB$_6$ is a paradigm mixed valent insulator [1] whose insulating gap is thought to arise from *f-d* hybridization [2,3]. The combination of optics [4] and transport [5] give evidence of a "small" transport gap ≈3-5 meV and also a "large" gap ≈10-20 meV. However the insulating state, whose transition starts below 50K, is not complete at low temperature (T), but instead saturates to a finite conductivity below T=4K implying some metallic states within the gap [1,6]. Renewed interest in this long-standing mystery has been generated by recent theoretical predictions of topologically protected surface states [7,8] and experimental evidence for robust surface conduction [9,10].

Angle-resolved photoemission spectroscopy (ARPES) is the most direct experimental method for observing such in-gap surface states and three recent ARPES papers [11-13] claim a possible observation of the predicted in-gap states corresponding to the topological theory. Two more recent ARPES papers claim instead a bulk origin of the X-point in-gap states [14] and observation of a non-topological polarity driven metallic surface state [15]. In this paper we first briefly review the prior photoemission leading up to these recent ARPES results using our own measurements. We then discuss the unresolved issues regarding detailed formation of the hybridization gap and summarize our efforts to probe the detailed *f*-band structure close to the Fermi energy ($E_F$) [16]. We then present examples of photon polarization and spatial dependence that enable a critical assessment of the newer ARPES interpretations [14,15] and provide further insight into the origin of the less-characterized Γ-point in-gap states.

*1.1 Experimental*

Angle-resolved photoemission (ARPES) measurements were performed at the MERLIN Beamline 4.0.3 at the Advanced Light Source (ALS) synchrotron in the photon energy (hv) range of 30 to 140 eV, employing an elliptically polarized undulator allowing selection of both linear and circular polarizations of the incident light. A Scienta R8000 electron energy analyzer with overall energy resolution as low as 7 meV was used with a liquid He flow cryostat 6-axis sample manipulator cooled to as low as 6K. Single-crystal samples of $SmB_6$, prepared from an aluminum flux, were cleaved and measured in ultra-high vacuum better than $8\times10^{-11}$ Torr. A refocused beam spot of $\approx 50\times20$ μm allowed spatial dependent characterization of the (001) cleaved surfaces. Above hv=40 eV the photoionization cross-section for Sm 4$f$ is very high relative to that for the Sm 5$d$ and B 2$p$ states and is further enhanced in the 4$d$→4$f$ resonance regime [17] above hv=130 eV.

The correspondence between hv and $k_z$ for the simple cubic Brillouin zone (BZ) of $SmB_6$ is provided in Fig. 1. The schematic illustrates the $k$-space coverage for a typical ±15 degree angular acceptance of the electron spectrometer and approximate high symmetry photon energies that pass through bulk Γ and X-points at normal emission. Also illustrated at the X-points are ellipses that approximate the size and anisotropy of bulk Sm 5$d$ electron pockets (section 2.3 below) as they approach the Fermi level ($E_F$).

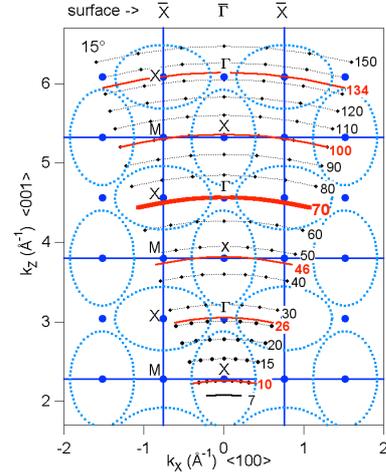

**Fig. 1.** (Color online) Schematic of the relation between the $SmB_6$ Brillouin zone and the Fermi-edge photoelectron excitation energy assuming an inner potential of $V_0$=14 eV. $k_z$-independent surface BZ labeling is also provided.

## 2. Basic Electronic Structure

*2.1 Mixed Valence*

$SmB_6$ is well-known to have mixed valence of ≈ 2.5+ corresponding to an average $f$-occupation $n_f \approx 5.5$ as first determined from magnetic susceptibility [18] and x-ray photoelectron spectroscopy [19], and more precisely quantified by x-ray absorption spectroscopy [20]. $SmB_6$ photoemission spectra for hv>40 eV are dominated by the essentially $k$-independent Sm 4$f$ emission, as shown in Fig. 2 for 140 eV, having the detailed fingerprint spectral structure of, respectively, 4$f^5$ and 4$f^4$ atomic multiplets originating from removing an electron from the 4$f^6$ and 4$f^5$ configurations of the $Sm^{2+}$ and $Sm^{3+}$ states that characterize the mixed valence. The $Sm^{2+}$ states lie within 4 eV of $E_F$ and the $Sm^{3+}$ states lie at more than 6 eV binding energy, reflecting the strong $f$-shell Coulomb repulsion that separates the valence states energetically and creates the strong correlations.

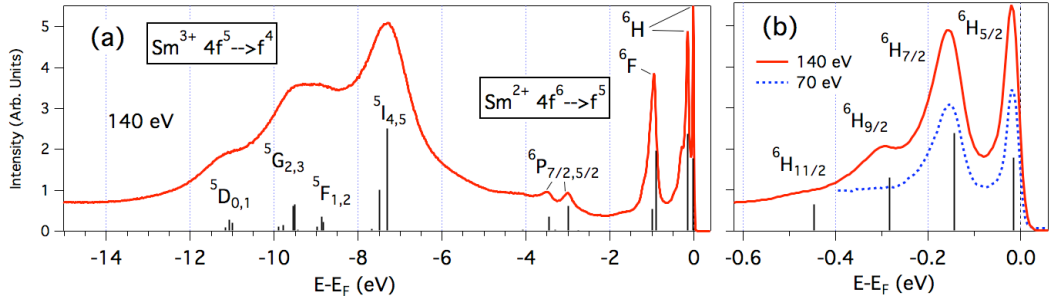

**Fig. 2.** (Color online) **(a)** Wide energy hν=140 eV angle-integrated spectrum of mixed-valent SmB$_6$ with comparison to final state multiplet energies for both Sm$^{3+}$ and Sm$^{2+}$ initial states. **(b)** Zoom on the $^6$H Sm$^{2+}$ final state multiplets with the lowest energy $f$-state gapped by ~15 meV below $E_F$.

Previous multiplet comparisons for SmB$_6$ [19,21] have been only to lower resolution XPS spectra. A resonant energy of 140 eV in Fig. 2 was selected in order to preferentially enhance the Sm$^{3+}$ multiplet states [17]. The final state multiplets, calculated with intermediate coupling [22], are appropriately scaled and offset in energy to best match the Sm$^{2+}$ and Sm$^{3+}$ manifolds. The improved experimental resolution allows finer details to be compared including (i) an intensity shoulder corresponding to Sm$^{3+}$ $^5$F states, (ii) the splitting of the Sm$^{2+}$ $^6$P$_{7/2}$ and $^6$P$_{5/2}$ states, and (iii) peaks corresponding to Sm$^{2+}$ $^6$H$_{9/2}$ and $^6$H$_{11/2}$ multiplet energies [23] that are observable in Fig. 2(b) only in the 4d→4f resonant energy range and not at lower non-resonant photon energies.

The strong mixed valence in SmB$_6$ is notably far from the standard Kondo regime of nearly trivalent cerium compounds, and the full role of the large fractional valence for the low energy transport properties has not been clearly elucidated. However, one key low energy scale property that reflects the hybridization gap is the finite ≈15 meV binding energy of the Sm$^{2+}$ $^6$H$_{5/2}$ multiplet peak, first observed in 2000 [24].

*2.2 Temperature dependence of Sm 4f states*

After the initial identification of the ≈15 meV $f$-gap, the next important experimental photoemission measurement was the temperature dependence of the 4f states and filling of the gap. Spatial inhomogeneity and thermal drifts of the sample position limited our early efforts to achieve a reliable Sm 4f T-dependence series from the (001) cleaved surface and hence initial measurements of the 4f T dependence used He I excitation of scraped surfaces [25,26]. More recently a very similar Sm 4f T-dependence has been measured on a cleaved surface using hν=8.4 eV excitation [27]. In contrast to these lower hν measurements where the maximal $f$-peak amplitude is less than 70% greater than that of the high binding energy $d$-state background, the high $f$-state cross-section at hν=70 eV allows the unprecedented observation of the full k-resolved T-dependent $f$-state coherence change with a nearly negligible background contribution as shown for the Γ-point in Fig. 3.

A key result of this and the earlier 4f T-dependence studies is that the 4f amplitude decline and $E_F$ gap filling associated with the 4f state broadening points to a hybridization coherence temperature scale of T*≈140K that is more like the "large" gap than the "small" gap inferred from transport, and the gradual changes of the Γ-point 4*f* T-dependent profiles in Fig. 3(b) exhibit no clear direct relation to the rapid resistivity rise below 50K. An understanding of the origins of these multiple energy scales comes from the T-dependence of the X-point band structure, presented in a separate paper [28].

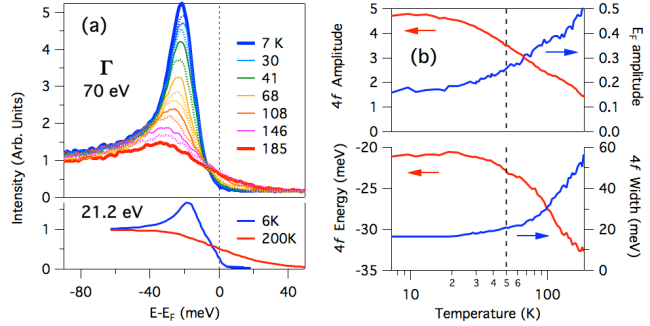

**Fig. 3.** (Color online) (a) Temperature dependence of the Sm 4*f*-states at the hν=70 eV normal emission Γ-point with comparison of the low and high temperature 4*f* spectra at 21.2 eV spectra of Ref. [26] (b) Extracted 4*f* peak and $E_F$ T-dependent profiles.

*2.3 k-resolved Valence Band Structure*

The basic ingredients of the near $E_F$ k-resolved valence band structure of SmB$_6$, i.e. LaB$_6$-like X-point 5*d* electron pockets hybridizing with narrow 5*f* states close to $E_F$, was also first illustrated in 2000 using hν=70 eV to probe the high symmetry Γ-X direction [24]. A progressive zoom of the k-resolved valence band structure for this same X-Γ-X cut at 70 eV is shown in Fig. 4 with comparison to theoretical bands from LaB$_6$ [29] scaled and offset to best match the experiment. In the wide energy spectrum in Fig. 4(a) dispersing bands of primarily B 2*p* character are visible between 11 eV and 2 eV binding energy in addition to the k-independent streaks of both the Sm$^{3+}$ and Sm$^{2+}$ 4*f* final state multiplets identified in Fig. 3. A 4 eV zoom of the Sm$^{2+}$ multiplets in Fig. 4(b) also highlights the Sm 5*d* X-point electron pockets exhibiting a 1.8 eV band minimum and in theory having nearly pure $d_{y^2-z^2}$ orbital character along $k_x$. A further zoom to 0.3 eV in Fig. 4(c) shows the appearance of strong k-variations of the Sm$^{2+}$ *f*-state intensities related to the 5*d* bands which exhibit a change in velocity along Γ-X from ≈5 eV-Å (m*=0.5) to ≈2 eV-Å (m*=2) from below to above the $^6H_{7/2}$ *f*-multiplet at 150 meV. A final zoom to 90 meV in Fig. 4(d) shows only the lowest energy $^6H_{5/2}$ multiplet with a dispersion minimum at Γ, and distinct in-gap states centered on the X-points and dispersing to $E_F$ with band velocity further reduced to 0.24 eV-Å (m*≈13). Additional in-gap intensity impinging on $E_F$ at the Γ-point is discussed later in section 3.2.

Angle-dependent ($k_x$-$k_y$) maps of the electronic structure in the high symmetry Γ-plane, as shown in Fig. 5(a) for hν=70 eV, reveal that both the in-gap $E_F$ states and the bulk 5*d*-states at -70 meV form elliptical contours centered on the X-points but with significantly different sizes. The in-gap ellipse has major and minor $E_F$ radii of $k_F$ = (0.41, 0.28) Å$^{-1}$ = (0.54, 0.37)×(π/a) with an area of ≈16% of the surface BZ $(2π/a)^2$, where π/a=0.76 Å$^{-1}$. In contrast the bulk 5*d* ellipse at -50 meV has a size of $k_F$ = (0.53, 0.41) Å$^{-1}$ = (0.70, 0.54)×(π/a) with almost double the cross-sectional area and 3D ellipsoidal volume of ≈8% of the bulk BZ. The total enclosed volume for three X-point ellipsoids of ≈24% is consistent with the mixed valence of ≈2.5 where 50% of the BZ corresponds to the trivalent LaB$_6$ Fermi surface volume.

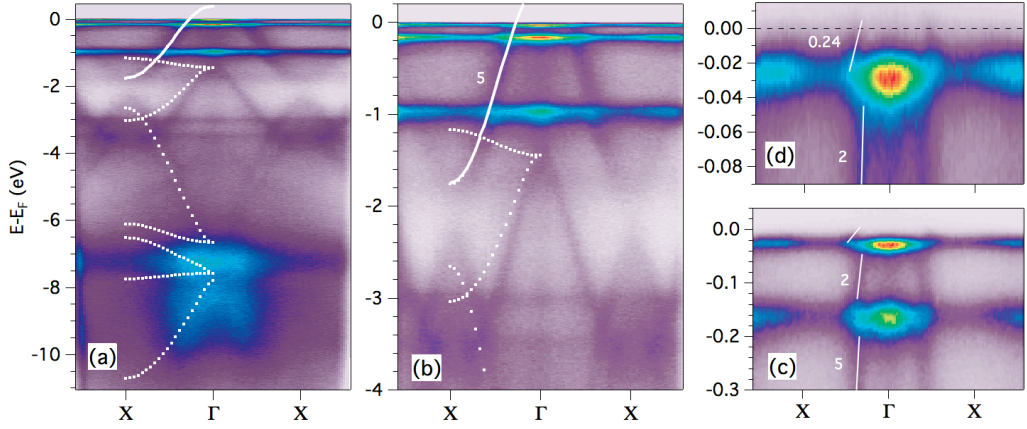

**Fig. 4.** (Color online) (a-d) Progressive zoom of the X-Γ-X valence band spectra measured at hν=70 eV using a sum of LH and LV photon polarizations and with comparison to theoretical B 2p (dotted) and 5d (solid) bands from LaB$_6$. X-point d-band velocity values in units of eV-Å are also labeled.

The distinct X-centered electron-like bands spanning the *f*-gap have been reported by three ARPES groups recently [11,13,14] within the energy range of hν=19-75 eV. At lower hν, a hint of *k*-dependent in-gap states was first reported for hν=10 eV excitation [30], and a locus of elliptical in-gap spectral weight was observed in a 7 eV laser-excited ARPES study [12]. Evidence for the 2D surface state nature of the in-gap states was provided by hν-dependent measurements from three of the ARPES studies [11-13] that observed a lack of $k_z$-dispersion of the $k_F$ value and of the band velocity $v_F$. However a fourth study [14] argued these states to be 3D bulk-like states with $k_z$-dependence.

Complementary to the hν-dependent mapping is the comparison of angle-dependent maps at discrete photon energies with different $k_z$ values to test whether the in-gap states possess the same size $E_F$ contours at the same surface BZ location while the bulk *d*-states are changing with $k_z$. A recent ARPES study has made such a comparison of hν=70 eV and hν=34 eV for $E_F$ and -350 meV angle-dependent maps and an apparent *lack* of X̄ elliptical contours in the hν=34 eV data was taken as key evidence for sensitivity to $k_z$ and a 3D character of the in-gap states [14]. Figure 5(b) repeats this hν=34 eV angle mapping for both LH and LV polarizations. We note from Fig. 1 that the $k_z$ associated with hν=34 eV lies midway between a Γ-plane and an X-plane of the simple cubic BZ and thereby spectra are likely to show bulk X-point ellipsoid features from both high-symmetry planes. Indeed this is what is observed for the -350 meV energy slice, but with a distinct polarization selectivity of the four Γ-plane $d_{y^2-z^2}$ and $d_{z^2-x^2}$ horizontal electron pockets for LH polarization and the single X-plane $d_{x^2-y^2}$ circular contour at normal emission for LV polarization. As expected, both the LH and LV -350 meV maps at hν=34 eV are significantly different from the high symmetry 70 eV map in Fig. 5(a), whereas the hν=34 eV Fermi-edge map for LH polarization shows distinct X̄ elliptical contours with the same size as that of the in-gap states at hν=70 eV. In contrast, the LV polarization hν=34 eV map, consistent with the data presented in

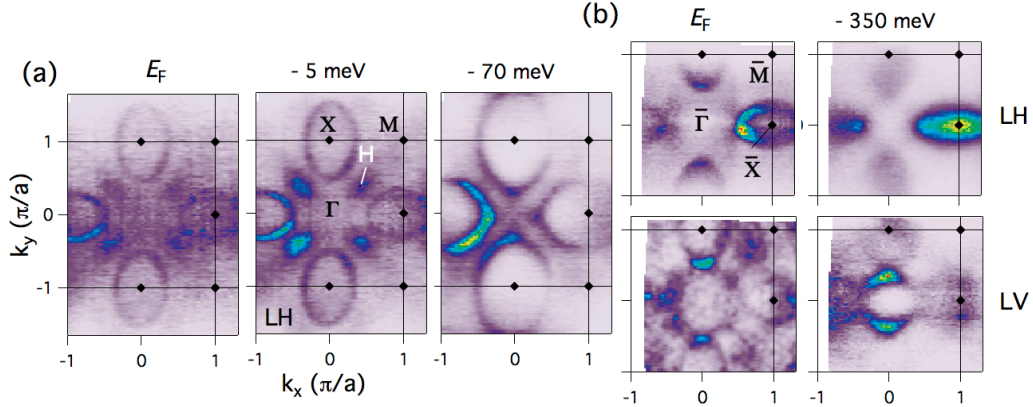

**Fig. 5.** (Color online) Constant binding energy slices of angular-dependent maps measured at (a) hν=70 eV and (b) hν=34 eV using LH and LV polarization as labeled. In-gap states are probed at $E_F$ and bulk $d$-states at -70 meV and -350 meV.

Ref. [14], creates the illusion of $k_z$-dependence by partial suppression of all but the tips of the $\bar{X}$ ellipses along $k_y$ only. This example highlights the need to have a measurement of both polarizations in order to avoid making an incorrect [14] interpretation of the data.

*2.4 4f band dispersion and gap formation*

Coming back to the $E_F$ map image in Fig. 5(a), additional spectral weight is observed along Γ-M that results from the spectral tail to $E_F$ of the dispersing $^6H_{5/2}$ multiplet. This high energy $f$-dispersion point, enhanced in the -5 meV map image and labeled $H$, is coincident with the near touching point of the neighboring 5$d$-band ellipses in the -70 meV map image. High resolution cuts through this point at hν=70 eV [16] reveal a hole-like $f$-dispersion whose peak centroid binding energy of ≈14 meV is distinctly smaller than the 21 meV minimum at the Γ-point and the 18 meV average at other $k$-points. Previously, only the 3 meV $f$-dispersion around the Γ-point maximum had been observed, using He I (21.2 eV) excitation [26]. We note that a recent He I study [15] claims to identify an electron pocket with $E_F$ crossing points midway along Γ-M as a metallic surface state originating from a boron-terminated surface. This feature is in fact coincident with the $H$-point which we clearly resolve as not crossing $E_F$ [16].

Key features of density functional (DFT) band calculations [31] include multiple $f$-sub-bands with degeneracy at the Γ-point, one $f$-sub-band that propagates above $E_F$ at the X-point and a 20 meV hybridization gap involving this $f$-sub-band along Γ-X (and >40 meV along X-M). While this uncorrelated 20 meV DFT gap has been cited numerous times in the literature as being in agreement with the "large" gap [12,31], we note that a 10X energy renormalization is required to align the least dispersive $f$-sub-band to experiment, which reduces this gap to 2 meV in closer agreement to the "small" transport activation gap. Also the $f$-sub-band splittings are observed to be 3-4X too large compared to the ARPES $f$-band width even with the 10X energy renormalization. Hence it is clear that the energy-renormalized DFT as well as recent correlated band calculations [32] is insufficient to describe both the narrow experimental $f$-band dispersion (7 meV) and the large $f$-gap (14-21 meV relative to $E_F$).

One can try to correct for the narrower $f$-band width by renormalizing the DFT theory even further and then energy shifting the chemical potential (of either the theory or experiment) to match the $f$-band central binding energy. This of course implies yet an even smaller hybridization gap that is yet further from the "large" transport gap, e.g. the 20 meV experimental tunneling gap [33,34]. Nevertheless, this scenario of a tiny hybridization gap has recently been proposed [14] as part of a model in which the (001) surface $E_F$ is pinned high in the conduction band and the "in-gap" states are actually of bulk conduction band origin. A more promising way to correct for the DFT disagreement with ARPES is to increase the renormalized $f$-$d$ hybridization strength which in addition to creating a larger gap has the beneficial effect of narrowing the occupied $f$-band width [16].

It is also important experimentally to look for the existence of the multiple theoretical $f$-sub-bands in order to go beyond a single $f$-peak dispersion analysis. As detailed elsewhere [16], quantification of the differences in the $f$-peak width and energy centroid between LH and LV polarization provide direct evidence for multiple sub-band contributions. Temperature dependent changes in the electronic structure at the X-point [28] provide further direct evidence for the X-point conduction band $f$-state, thus ruling out the model [14] of a surface $E_F$ being pinned entirely above the bulk X-point $f$-dispersion that hides a tiny gap.

## 3. Cleave Surface Inhomogeneity

### 3.1 Cleave surface spatial dependence

The ionic character of the Sm-B$_6$ bonding implies the possibility of different surface terminations having opposite polarity, which could result in different surface band bendings and also drive surface reconstructions. Indeed recent STM measurements [34,35] observe numerous types of domains on the (001) cleaved surface including unreconstructed 1×1 Sm or B terminations, an ordered 2×1 Sm reconstruction, and disordered Sm-chain reconstructions as well as just disordered boron termination with majority existence of the disordered reconstructions that are likely consequences of the unstable polar surfaces. Here we use our small beam spot of <50 μm to probe the spatial dependence of ARPES spectra across a single cleaved surface and find distinct differences on this macroscopic length scale. Fig. 6(a) compares 5 eV wide valence band spectra of two points on a cleaved surface, labeled "A" and "B", that exhibit complementary contrast in the visibility of boron $2p$ valence bands and Sm $4f$ intensity. Sample optimization based on absolute photoemission signal near $E_F$ would naturally select the "A" surface region. However it is revealed in Fig. 6(a) that the "A" point also contains additional $k$-independent streaks of intensity at 0.7, 1.5 and 4 eV binding energy that are not present at the "B" point. The relative spacing of these extra peaks indicates that they originate from broadened and shifted versions of the Sm$^{2+}$ final state multiplets, as is well known for rare earth surface atoms. Furthermore the energy shift of the broadened peaks is observed to vary on the surface with the least binding energy surface Sm multiplet peak varying between 0.4 to 0.7 eV [36]. This is suggestive of a variation of Sm surface sites ranging from step edges to uniform or reconstructed terraces as observed by STM. Such shifted peaks were observed in early studies of polycrystalline SmB$_6$ [17] cleaved *in situ* by argon ion bombardment.

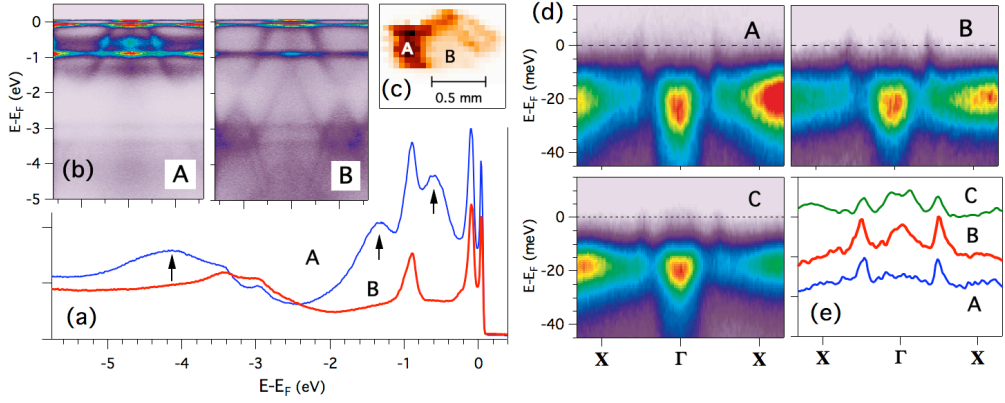

**Fig. 6.** (Color online) **(a)** Angle-integrated and **(b)** X-Γ-X hν=70 eV valence spectra of a two regions (A,B) of a cleaved surface of SmB$_6$ exhibiting distinct differences in extra broad Sm 4$f$ peaks (A) and enhanced visibility of B 2$p$ bands (B). **(c)** In-situ XY scan of the 0.5 eV photoelectron intensity identifying the locations of points A and B. **(d)** LV polarization near-$E_F$ X-Γ-X spectra and **(e)** Fermi-edge intensity profiles from the A and B points and a third surface type (C) all exhibiting a different presence of Γ-point in-gap states.

A recent He I excitation study of cleaved SmB$_6$ [15] has observed a single broad peak centered at 0.8 eV (with larger width than in Fig. 6) which then exhibits a rapid intensity decay within hours after cleaving a fresh surface. Similarly we also observed significant intensity reduction of this peak within hours after a cleave, but with a rate of decay that is strongly dependent on the initial binding energy of the broadened peaks [36]. The He I study identified the energy of the time-dependent 0.8 eV peak as a statistical average of ≈2 eV and ≈0 eV surface states derived from boron dangling bonds on the Sm- and B-terminated surfaces, respectively, whereas we identify it as a 4$f$-peak originating from non-bulk-coordinated Sm surface atoms.

*3.2 Γ-point in-gap state variation*

Another important difference between surface regions is the presence or absence of the Γ-point in-gap state as shown in Fig. 6(b). While the "B" surface exhibits a fuzzy presence of the Γ-point states (for LV polarization only), the "A" region is distinctly missing any in-gap Γ-point states (for both LH and LV polarization). This suggests that the fuzzy Γ-point surface state is primarily associated with the B-terminated surface. Furthermore, yet a third type of surface was observed, labeled "C" in Fig. 6(c), that exhibits a distinct two-peak structure of Γ-point states with a larger ~0.09 Å$^{-1}$ $k_F$ separation and very steep, light effective mass (<0.1 m$_e$), dispersion. Overall this second new type of Γ-point in-gap state, not reported previously, appears most often on "aged" surface regions that initially exhibited the surface Sm atom signature, but also with variability of the band intensity close to $E_F$ relative to the intensity of the X-point in-gap state dispersion. While the $k_F$ separation of the two in-gap bands is significantly larger than recently observed dHvA tiny orbit size ($k_F$ = 0.038 Å$^{-1}$), the light effective mass is comparable to the dHvA value of 0.074 [37]. A more detailed exposition of the hν-dependence, angle-dependence, K-dosing and other characteristics of these Γ-point in-gap states will be presented elsewhere [36]. The importance of clarifying these normal emission in-gap state variations lies in the fact that the occurrence of a strong topological insulator requires [32,38-40] an odd number of Dirac points per surface BZ. Since by symmetry there are two $\bar{X}$ points per zone but only one $\bar{\Gamma}$ point, the latter is crucial for giving an odd number. Nonetheless, the additional presence of ordinary surface states that might occur only for some specific surface, such as the recently observed ≈2 eV surface state derived from the Sm-terminated (001) surface [15], is not precluded.

## 4. Conclusion

Early photoemission of the mixed valence spectra, the Sm 4$f$ gap and T-dependence, and basic near-$E_F$ electronic structure of SmB$_6$ has been reviewed and revisited with newer measurements that reveal finer details of final state multiplets, dramatically larger 4$f$ amplitude variation, 7 meV dispersion of the $f$-bands, and distinct X-point bands spanning the 14-21 meV 4$f$ gap up to $E_F$. The assignment of these recent multiply reported X-point states as being 2D in-gap surface states is confirmed and experimental contradictions with alternate interpretations of surface-$E_F$ pinning and polarity-driven surface states are pointed out. Linear polarization dependence and small-spot spatial dependence, not previously employed in the recent SmB$_6$ ARPES studies, are shown to be crucial to the full ARPES characterization. Results of many more details to be published elsewhere are summarized including elucidation of the $f$-sub-band gap formation and observation of multiple Γ-point in-gap states.

## Acknowledgment

Supported by U.S. DOE at the Advanced Light Source (DE-AC02-05CH11231), and at U. Michigan (DE-FG02-07ER46379). KS was supported in part by NSF under Grant No. ECCS-1307744 and the MCubed program at U. Michigan. B.I.M. was supported by the NRF (Grant No.2009-0079947). J.S.K. was supported by the NRF (Grant No. 2011-0022444).